# An ecologically valid examination of event-based and time-based prospective memory using immersive virtual reality: the influence of attention, memory, and executive function processes on real-world prospective memory.

Panagiotis Kourtesis[a,b,c,d,e,f,g,h]* and Sarah E. MacPherson[a,b]

[a]*Human Cognitive Neuroscience, Department of Psychology, University of Edinburgh, Edinburgh, UK;*

[b]*Department of Psychology, University of Edinburgh, Edinburgh, UK;*

[c]*Lab of Experimental Psychology, Suor Orsola Benincasa University of Naples, Naples, Italy;*

[d]*Interdepartmental Centre for Planning and Research "Scienza Nuova", Suor Orsola Benincasa University of Naples, Naples, Italy;*

[e]*National Research Institute of Computer Science and Automation, INRIA, Rennes, France;*

[f]*Univ Rennes, Rennes, France;*

[g]*Research Institute of Computer Science and Random Systems, IRISA, Rennes, France;*

[h]*French National Centre for Scientific Research, CNRS, Rennes, France.*

\* Panagiotis Kourtesis, Department of Psychology, University of Edinburgh, 7 George Square, Edinburgh, EH8 9JZ, United Kingdom. Email: pkourtes@exseed.ed.ac.uk

# Abstract

Studies on prospective memory (PM) predominantly assess either event- or time-based PM by implementing non-ecological laboratory-based tasks. The results deriving from these paradigms have provided findings that are discrepant with ecologically valid research paradigms that converge on the complexity and cognitive demands of everyday tasks. The Virtual Reality Everyday Assessment Lab (VR-EAL), an immersive virtual reality (VR) neuropsychological battery with enhanced ecological validity, was implemented to assess everyday event- and time-based PM, as well as the influence of other cognitive functions on everyday PM functioning. The results demonstrated the importance of delayed recognition, planning, and visuospatial attention on everyday PM. Delayed recognition and planning ability were found to be central in event- and time-based PM respectively. In order of importance, delayed recognition, visuospatial attention speed, and planning ability were found to be involved in event-based PM functioning. Comparably, planning, visuospatial attention accuracy, delayed recognition, and multitasking/task-shifting ability were found to be involved in time-based PM functioning. These findings further suggest the importance of ecological validity in the study of PM, which may be achieved using immersive VR paradigms.

Keywords: prospective memory; episodic memory; attention; executive functions; ecological validity

# Introduction

Prospective memory (PM) involves the ability to remember to initiate an action in the future. The PM action may be associated with either a certain event (i.e., event-based PM; e.g., when you pass the supermarket, buy a pint of milk) or a specific time (time-based PM; e.g., at 3 pm watch the weather forecast). PM functioning is crucial for many aspects of everyday life, varying from minor (e.g., brushing your teeth after having breakfast) to vital importance (e.g., having an insulin injection after dinner). PM impairment is common in healthy older individuals (Huppert, Johnson, & Nickson, 2000), as well as in clinical populations such as Parkinson's disease (Kliegel, Altgassen, Hering, & Rose, 2011), individuals with mild cognitive impairment (Costa, Caltagirone, & Carlesimo, 2011), Alzeimer's disease (Jones, Livner, & Bäckman, 2006), and autism spectrum disorder (Sheppard, Bruineberg, Kretschmer-Trendowicz, & Altgassen, 2018). The study of everyday PM may assist in our understanding of real-life PM functioning, and consequently provide ways to improve functioning in individuals with PM impairments.

## *Cognitive functions associated with prospective memory*

The two main theoretical models of PM are the preparatory attentional and memory (PAM) theory and the multiprocess theory (Anderson, McDaniel, & Einstein, 2017). These two theoretical models have attempted to explain the implications of attentional and memory processes in everyday PM functioning. Based on the PAM theory, PM functioning relies on attentional processes for monitoring both environmental and internal cues, which may assist an individual in recalling their intended PM action (Smith, 2003; Smith, Hunt, McVay, & McConnell, 2007). The multiprocess framework suggests that passive monitoring may also assist in detecting target cues, resulting in reflexive associative retrieval of the intended PM action (McDaniel & Einstein, 2000; McDaniel & Einstein, 2007). More recently, the dynamic multiprocess framework has been proposed, which suggests a dynamic interplay between top-down and bottom-up monitoring and retrieval processes for PM functioning (Scullin, McDaniel, & Shelton, 2013; Shelton & Scullin, 2017). The dynamic multiprocess framework emphasises the importance of contextual information and/or formulating an effective plan of action in PM functioning (Scullin *et al.*, 2013; Shelton & Scullin, 2017). Contextual information may indicate the probability of detecting a target cue related to the intended PM action in the current environment (Marsh, Hicks, & Cook, 2008; Scullin *et al.*, 2013; Shelton & Scullin, 2017). In addition, an effective plan of action is facilitated by associating potential environmental cues with the intended PM action (Scullin *et al.*, 2013; Shelton & Scullin, 2017).

The formulation of an effective plan of action is based on Gollwitzer's *implementation intention*, which is an encoding technique for future actions by associating a cue with the intended action (i.e., "if I encounter X then I will do Y"; Gollwitzer, 1999). Planning ability, an executive function, involves thinking about the future, organizing and prioritizing future actions (Miyake *et al.*, 2000; Morris & Ward, 2004). In everyday life, planning defines when and where an action will take place, and updates/prioritises the plan of action based on the received information (e.g., I received a letter notifying me of my overdue electricity bill, so, tomorrow I will need pay it after work; Morris & Ward, 2004). Indeed, effective planning has been found to improve PM functioning (Azzopardi, Auffray, & Kermarrec, 2017; Gonneaud *et al.*, 2011; Liu & Park, 2004; McDaniel & Scullin, 2010; McFarland & Glisky, 2009; Milne, Orbell, & Sheeran, 2002; Mioni & Stablum, 2014; Vanneste, Baudouin, Bouazzaoui, & Taconnat, 2016; Zuber, Mahy, & Kliegel, 2019). Another executive function which is often found to contribute to PM functioning is task-shifting (Azzopardi *et al.*, 2017; Gonneaud *et al.*, 2011; McFarland & Glisky, 2009; Mioni & Stablum, 2014; Vanneste *et al.*, 2016; Zuber *et al.*, 2019). Task-shifting pertains to the ability to inhibit one's attention on the current task, and shift it to another task (Miyake *et al.*, 2000).

However, as mentioned above, everyday PM consists of both event- and time-based PM (Einstein & McDaniel, 1996). Hence, including only event-based tasks in the assessment of PM does not allow for generalisation of the results to PM functioning in the daily life (e.g., Einstein, Smith, McDaniel, & Shaw, 1997; Einstein, McDaniel, Manzi, Cochran, & Baker, 2000; Mullet, Scullin, Hess, Scullin, Arnold, & Einstein, 2013; Scullin, McDaniel, Shelton, & Lee, 2010; Smith, 2003; Smith *et al.*, 2007). Time-based tasks may have similar aspects to event-based tasks (e.g., a cue which reminds an individual of the intended PM action), but they significantly differ in other aspects (Glickson, & Myslobodsky, 2006; Zuber *et al.*, 2019). A key difference is the requirement to monitor the time in time-based tasks (Glickson, & Myslobodsky, 2006; Zuber *et al.*, 2019). However, individuals do not appear to constantly monitor time. Instead, the most common behaviour is to monitor time strategically (Cona, Arcara, Tarantino, & Bisiacchi, 2012; Kliegel, Martin, McDaniel, & Einstein, 2001). By implementing strategic time monitoring, individuals check the time and then estimate when they should check it again, and adjust correspondingly the frequency of their monitoring (Cona *et al.*, 2012; Kliegel *et al.*, 2001).

### *The importance of ecological validity*
Evidence for the role of attentional, memory, and executive functioning processes in PM functioning derives predominantly from experiments which have implemented laboratory tasks. These laboratory tasks incorporate simple, static stimuli within a highly controlled environment, which do not resemble the complexity of real-life situations (Parsons, 2015). For example, a PM laboratory experiment may incorporate a single on-going task such as a computerised lexical decision task where the participant should indicate whether the presented word is a real or fictional word, together with a comparable computerised PM task, where the participant should indicate when a specific word appears. Hence, the laboratory tasks incorporate a simple testing environment (e.g., two-dimensional) and stimuli (e.g., static), with almost automatic responses (i.e., in milliseconds) provided using a keyboard, a relatively short testing duration (e.g., 15 minutes) and delay from encoding the PM intention (e.g., up to 10 minutes), and a single on-going task and a PM task with low demands.

Importantly, the implementation of non-ecologically valid PM tasks may provide discrepant results compared to ecologically valid research paradigms (Marsh, Hicks, & Landau, 1998). Ecologically valid tasks substantially converge with the complexity and cognitive demands of corresponding everyday tasks (Chaytor & Schmitter-Edgecombe, 2003; Franzen & Wilhelm, 1996; Spooner & Pachana, 2006), which allows the findings to be generalised to everyday functioning (Chaytor & Schmitter-Edgecombe, 2003; Haines *et al.*,

2019; Higginson, Arnett, & Voss, 2000; Mlinac & Feng, 2016; Parsons, 2015; Phillips, Henry, & Martin, 2008). The implementation of ecologically valid tasks is therefore required to examine the role of attentional, memory, and executive functioning processes in everyday PM functioning.

*Naturalistic Experiments*

Ecologically valid research paradigms have predominantly involved assessments in real-life settings such as performing errands in a shopping centre or a pedestrianized street (e.g., Garden, Phillips, & MacPherson, 2001; Marsh *et al.*, 1998; Shallice & Burgess, 1991). In PM research, these naturalistic experiments do not tend to consider diverse types of PM tasks within the same paradigm. The most replicated discrepancy between the findings from laboratory and naturalistic tasks is the so-called age-prospective-memory-paradox (Kvavilashvili, Cockburn, & Kornbrot, 2013; Niedźwieńska & Barzykowski, 2012; Phillips *et al.,* 2008; Uttl, 2008). In laboratory tasks, there are significant age-related differences on both event-based and time-based tasks between younger and older adults, while in naturalistic tasks, older adults perform at least comparably to younger adults (Kvavilashvili *et al.*, 2013; Niedźwieńska & Barzykowski, 2012; Phillips *et al.,* 2008; Uttl, 2008). This paradox may be attributed to factors such as individual differences (Phillips *et al.,* 2008; Shelton & Scullin, 2017; Uttl, 2008) in the use of compensatory strategies and external memory aids, which are common strategies in daily life for performing PM tasks (Phillips *et al.,* 2008; Uttl, 2008).

However, real-world tasks suffer from several limitations such as an inability to be standardized for use in other clinics or laboratories, reduced feasibility with certain neurological populations (e.g., schizophrenia patients), augmented time, fiscal, procedural and bureaucratic demands, as well as diminished experimental control over external factors (Parsons, 2015). An example of diminished control can be seen in the study of Niedźwieńska and Barzykowski (2012) where the scoring was dichotomous in terms of 0 or 1, without considering other factors (e.g., using external aids) that assist in performing PM tasks, and this may have caused ceiling effects.

*Virtual Reality*

An alternative approach to achieve ecological validity is the implementation of non-immersive virtual environments, which are able to simulate real-life tasks, are cost-effective, require less administration time, enable increased experimental control, and can be implemented in other clinical or research settings (Parsons, McMahan, & Kane, 2018; Werner & Korczyn, 2012; Zygouris & Tsolaki, 2015). However, non-immersive virtual reality (VR) tasks can be difficult for individuals without gaming backgrounds (e.g., competency in using a keyboard and mouse to interact with the virtual environment; Zaidi, Duthie, Carr, & Maksoud, 2018) and they are substantially less ecologically valid than immersive VR tests (Parsons *et al.*, 2018; Rizzo, Schultheis, Kerns, & Mateer, 2004). On the other hand, immersive VR tests benefit from the same advantages as non-immersive ones, but they also benefit from enhanced ecologically validity (Bohil, Alicea, & Biocca, 2011; Parsons, 2015; Rizzo *et al.*, 2004; Teo *et al.*, 2016). Notably, the first-person view and ergonomic/naturalistic interactions within immersive VR environments are proximal to real-life actions, as a result they significantly alleviate the differences in performance between gamers and non-gamers (Zaidi *et al.*, 2018). Therefore, an immersive VR research paradigm appears be the most efficient approach to study everyday PM functioning.

*The Virtual Reality Everyday Assessment Lab*

The Virtual Reality Everyday Assessment Lab (VR-EAL) was recently developed as an immersive VR neuropsychological battery assessing everyday cognitive functions such as PM (i.e., event-based and time-based), episodic memory (i.e., immediate and delayed recognition), attentional processes (i.e., visual, visuospatial, and auditory), and executive functioning (i.e., planning and task-shifting abilities; Kourtesis, Korre, Collina, Doumas, & MacPherson, 2020b). The convergent, construct, and ecological validity of the VR-EAL has been examined against established ecologically valid paper-and-pencil tests including the Cambridge Prospective Memory Test (Wilson *et al.*, 2005), the Rivermead Behavioral Memory Test–III (Wilson, Cockburn, & Baddeley, 2008), the Test of Everyday Attention (Robertson, Ward, Ridgeway, & Nimmo-Smith, 1996), and the Behavioral Assessment of Dysexecutive Syndrome (Wilson, Alderman, Burgess, Emslie, & Evans, 1997; Kourtesis, Collina, Doumas, & MacPherson, 2020a). Importantly, there is no performance difference in the VR-EAL between gamers and non-gamers. Also, participants rated the VR-EAL scenarios and tasks as significantly more similar to everyday tasks than the equivalent ecologically valid paper-and-pencil tasks (Kourtesis *et al.*, 2020a).

Hence, the VR-EAL appears to better resemble the complexity and cognitive demands of real-life tasks, as well as assessing both event- and time-based PM, and cognitive processes (i.e., attentional, memory, and executive functioning) which have been found to be associated with PM functioning. The VR-EAL is significantly closer to real life situations due to attributes such as the complexity and synthesis of the environment (i.e., a realistic and complex 360º environment), the quality of the stimuli (i.e., dynamic and realistic), the duration of the experiment (i.e., approximately 70 minutes), the naturalistic and ergonomic interactions within the testing environment, the response type (i.e., naturalistic) and time (i.e., in seconds), the length of the delay between encoding the PM intention and performing the PM action (i.e., 15 – 60 minutes), as well as including several cognitively demanding ongoing and PM tasks.

*The current study*

To our knowledge, the present study is the first to explore the cognitive functions that are involved in everyday event- and time-based PM functioning using an ecologically valid immersive VR paradigm in a single sample. As discussed above, the relevant literature postulates that attentional, memory, and executive processes are implicated in PM functioning. Thus, we hypothesize that attention, memory, and/or executive functioning will be involved in everyday PM functioning. However, since ecologically valid and laboratory paradigms does not provide convergent findings, we expect that the implementation of an ecologically valid assessment may indicate the increased importance of certain cognitive functions (e.g., delayed recognition, visuospatial attention, and planning), while it may decrease the importance of other cognitive functions (e.g., immediate recognition, visual attention, and multitasking). In terms of delayed recognition, the VR-EAL scenario lasts approximately 70 minutes which allows one to investigate the importance of recognising a cue after a long delay. Also, the inclusion of several diverse everyday tasks (PM and non-PM) within a realistic scenario may indicate the importance of prioritising activities (i.e., planning ability) in real life PM. Furthermore, the VR-EAL activities are performed in a naturalistic way, which substantially differ from the almost immediate responses in laboratory paradigms

(i.e., reaction times in milliseconds). Hence, allowing the examinee to perform the tasks in a naturalistic way (e.g., in seconds) may reveal the decreased importance of multitasking/task-shifting ability in everyday PM functioning. Finally, the VR-EAL scenario takes place within a complex, dynamic, and immersive 360º environment, hence, the importance of visuospatial attention processes (i.e., directing attention to a specific area of the environment and identifying items of interest within this specific area) may be seen as more important than visual attention processes (i.e., identifying items of interest within the visual field).

## Methods

### *Participants and Procedure*

Participants were recruited via social media and the internal mailing list of the University of Edinburgh and were the same cohort as in Kourtesis *et al.* (2020a). Forty-one participants (21 females) with a mean age of 29.15 years (SD = 5.80, range = 18-45) and mean education of 13.80 years (SD = 2.36, range = 10-16) were recruited. The study was approved by the Philosophy, Psychology and Language Sciences Research Ethics Committee of the University of Edinburgh. Written informed consent was obtained from each participant. All participants received verbal and written instructions regarding the procedures, possible adverse effects of immersive VR (e.g., cybersickness), utilization of the data, and general aims of the study. The participants attended the session individually. Upon their arrival, they were instructed by the examiner how to use the VR equipment. Then the participants were immersed into the VR-EAL, where they had a practice period without time restrictions to get familiarised with the basic controls and configuration. When the participants felt comfortable interacting within the virtual environment of VR-EAL, the VR-EAL's scenario commenced. All participants completed all tasks (i.e., non-PM and PM) and the full VR-EAL scenario. After exposure to VR, the Virtual Reality Neuroscience Questionnaire (VRNQ; Kourtesis, Collina, Doumas, &MacPherson, 2019b) was administered to assess cybersickness symptomatology.

### *Materials*

### *Hardware*

An HTC Vive HMD with two lighthouse stations for motion tracking and two HTC Vive wands with six degrees of freedom (6DoF) for navigation and interactions within the virtual environment were implemented in accordance with our previously published technological recommendations for immersive VR research (Kourtesis, Collina, Doumas, & MacPherson, 2019a). The spatialized (bi-aural) audio was facilitated by a pair of Senhai Kotion Each G9000 headphones. The size of the VR area was 5m$^2$, which provides an adequate space for immersion and naturalistic interaction within virtual environments (Borrego, Latorre, Alcañiz, & Llorens, 2018). The HMD was connected to a laptop with an Intel Core i7 7700HQ 2.80GHz processor, 16 GB RAM, a 4095MB NVIDIA GeForce GTX 1070 graphics card, a 931 GB TOSHIBA MQ01ABD100 (SATA) hard disk, and Realtek High Definition Audio.

*VR-EAL*

VR-EAL assesses everyday cognitive functions such as PM, episodic memory (i.e., immediate and delayed recognition), executive functioning (i.e., planning, multitasking) and selective visual, visuospatial and auditory (bi-aural) attention within a realistic immersive VR scenario lasting around 70 minutes (Kourtesis *et al.*, 2020a). See Table 1 and Figures 1 and 2 for a summary of the VR-EAL tasks assessing each cognitive ability. Also, a brief video recording of the VR-EAL may be accessed at this hyperlink: https://www.youtube.com/watch?v=IHEIvS37Xy8&t .

Table 1. VR-EAL tasks and score ranges

| Scene | Cognitive Function | Task | Score |
|---|---|---|---|
| 3 | Prospective memory | Write down the notes for the errands. | 0 – 6 |
| 3 | Immediate recognition | Recognising items on the shopping list. | 0 – 20 |
| 3 | Planning | Drawing the route to be taken. | 0 – 19 |
| 6 | Multitasking | Cooking task (preparing breakfast). | 0 – 16 |
| 6 | Prospective memory – event based | Take medication after breakfast. | 0 – 6 |
| 8 | Selective visuospatial attention | Collect items from the living room. | 0 – 20 |
| 8 | Prospective memory – event based | Take the chocolate pie out of the oven. | 0 – 6 |
| 10 | Prospective memory – time based | Call Rose at 10 am. | 0 – 6 |
| 12 | Selective visual attention | Find posters on both sides of the road. | 0 – 16 |
| 14 | Delayed recognition | Recognising items from the shopping list. | 0 – 20 |
| 15 | Prospective memory – time based | Collect the carrot cake from the bakery at 12 pm. | 0 – 6 |
| 16 | Prospective memory – event based | False prompt before going to the library. | -6 – 0 |
| 17 | Prospective memory – event based | Return the red book to the library. | 0 – 6 |
| 19 | Selective auditory attention | Detect sounds from both sides of the road. | 0 – 32 |
| 20 | Prospective memory – time based | False prompt before going back home. | -6 – 0 |
| 21 | Prospective memory – event based | Back home, give the extra pair of keys to Alex. | 0 – 6 |
| 22 | Prospective memory – time based | Take the medication at 1pm. | 0 – 6 |

*The tasks are presented in the same order as they are performed within the scenario. The table derives from Kourtesis et al. (2020a)*

Figure 1. VR-EAL Storyline: Scenes 3 - 12.

**Scene 3**        **Scene 3**

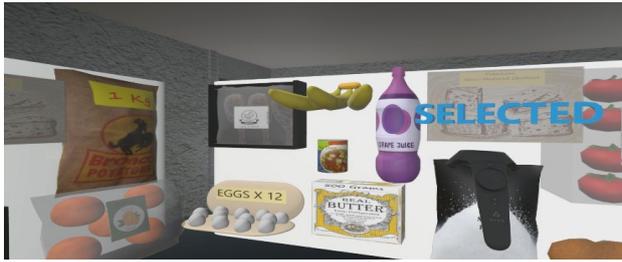 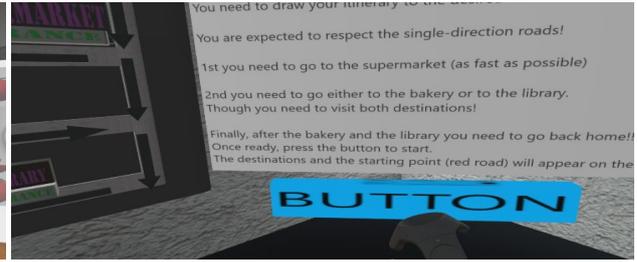

**Scene 6**        **Scene 6**

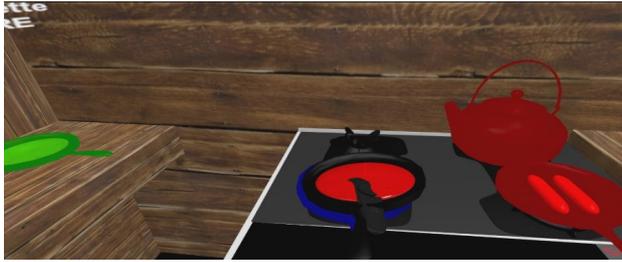 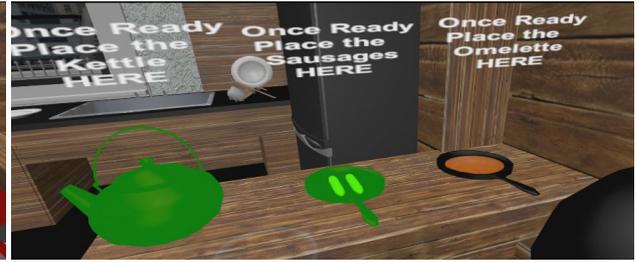

**Scene 6**        **Scene 8**

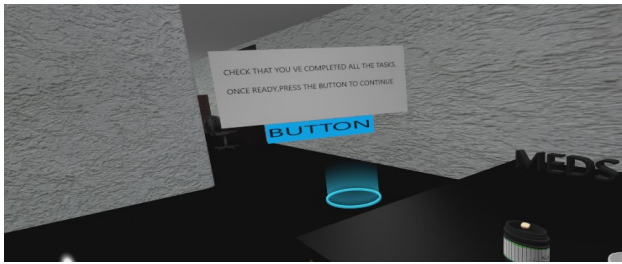 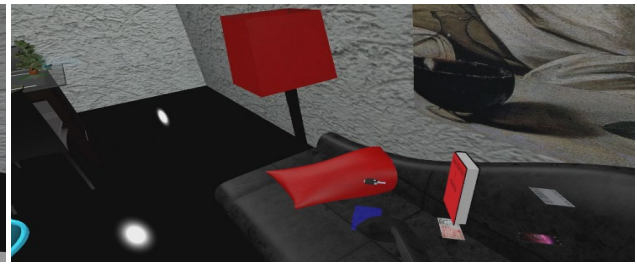

**Scene 8**        **Scene 10**

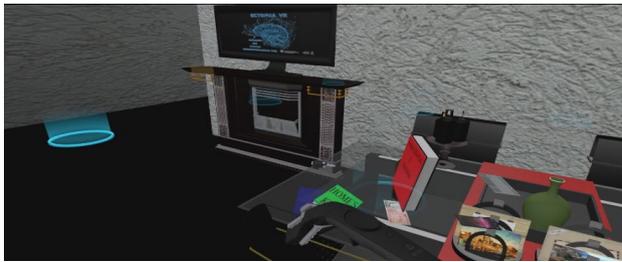 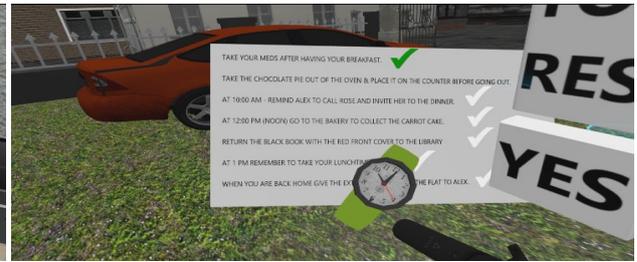

**Scene 12**        **Scene 12**

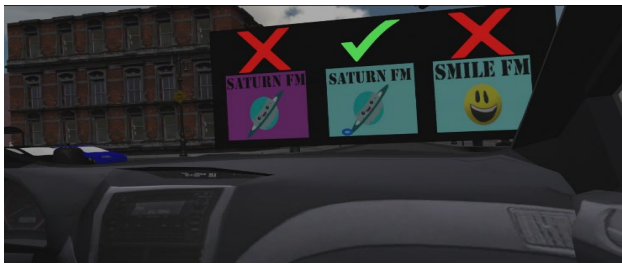 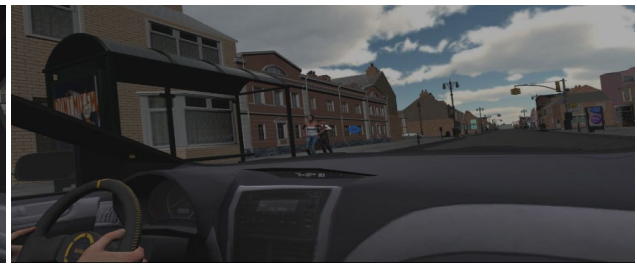

*Derived from Kourtesis et al. (2020b).*

Figure 2. VR-EAL Storyline: Scenes 14 – 22

**Scene 14**

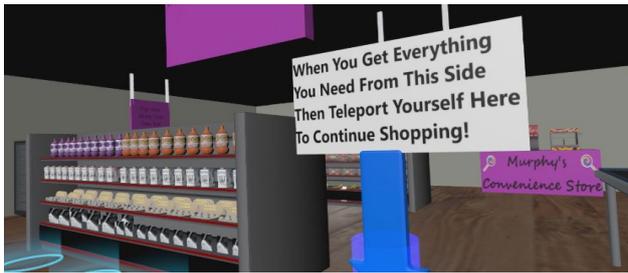

**Scene 14**

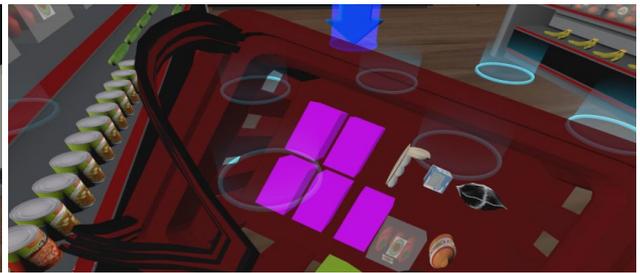

**Scene 15**

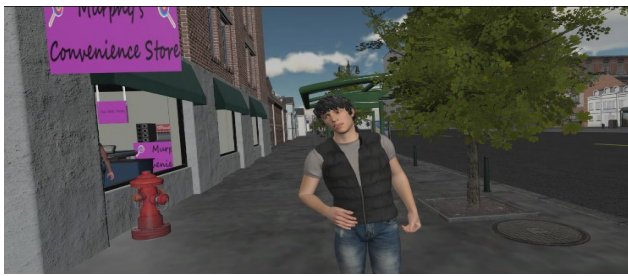

**Scene 17**

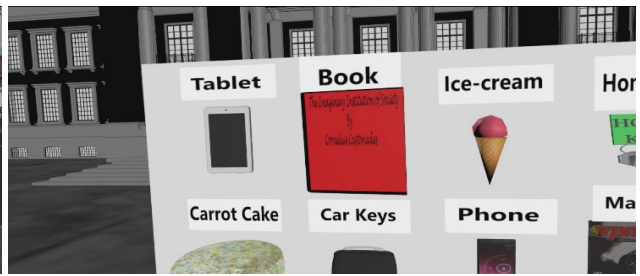

**Scene 19**

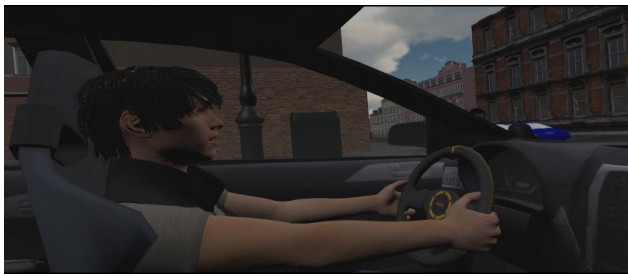

**Scene 19**

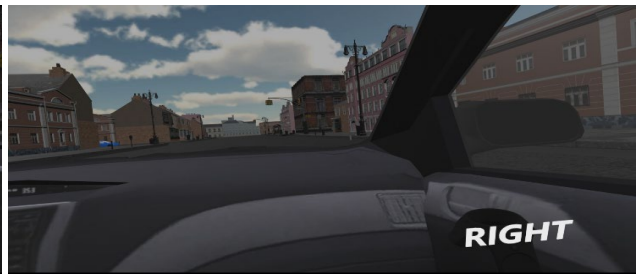

**Scene 20**

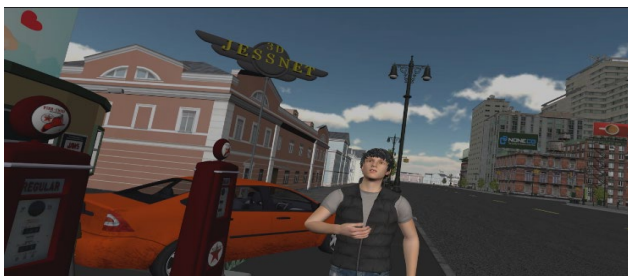

**Scene 22**

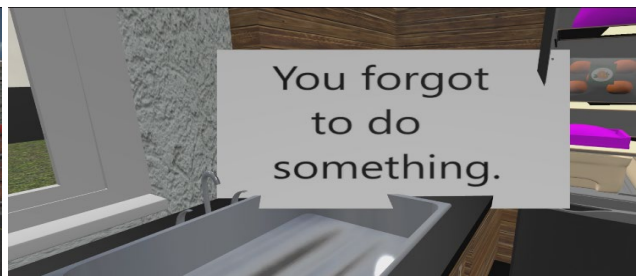

**Scene 22**

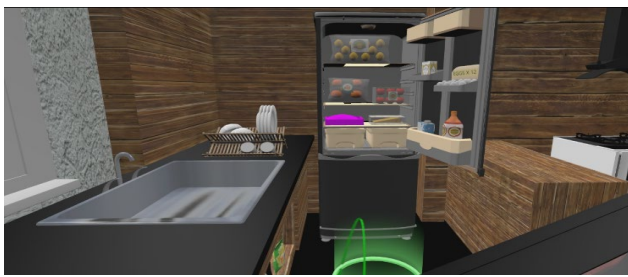

**Scene 22**

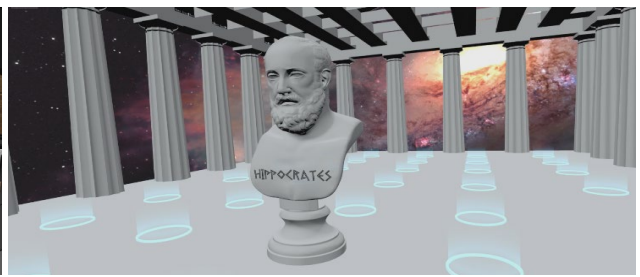

*Derived from Kourtesis et al. (2020b).*

*Prospective Memory*. VR-EAL considers both event-based and time-based PM tasks. The PM tasks are divided into five event-based tasks and four time-based tasks. In the event-based tasks, the participant should remember to perform a PM action when a specific event occurs (e.g., take medicines after breakfast). In the time-based tasks, the examinee should remember to perform a planned action at a specific time (e.g., call Rose at 12 pm). The following is a scoring example. At the end of a scene, the examinees should press a button to confirm that all the tasks in the scene have been completed. If the examinees have already taken their medication (i.e., PM task) before pressing the final button, then the scene ends, and the examinees receive 6 points. Otherwise, the first prompt appears (i.e., "You Have to Do Something Else"). If the examinees then follow the prompt and take their medication, they receive 4 points. If the examinees press the final button again, then the second prompt appears (i.e., "You Have to Do Something After Having your Breakfast"). If the examinees follow this prompt and take their medication, they receive 2 points. If the examinees press the final button again, then the third prompt appears (i.e., "You Have to Take Your Meds"). If the examinees follow this prompt and take their medication, they then receive 1 point. If the examinees repress the final button without ever taking their medication, they get zero points, and the scene ends. Scores range from 0 to 6.

There are also two misleading tasks. One is related to an event (i.e., event-based) and the other to a specific time (i.e., time-based). However, in the misleading tasks, the examinee should respond negatively to the virtual human prompts to perform a task because there is not a task to perform. The misleading tasks serve two purposes. The first is deception to examine whether the examinee is simply responding affirmatively to all PM task prompts. If the examinee responds affirmatively, then s/he loses points (see Table 1). The second is that these false prompts are common in everyday life (e.g., a friend invites you to watch an episode of a TV series together, but you decline because you remember that the episode will be aired tomorrow and not today). Hence, their inclusion increases the ecological validity of VR-EAL.

*Episodic Memory*. Both immediate and delayed episodic memory are assessed. Firstly, the participant needs to memorize a shopping list which is presented audio-visually. Immediately after the presentation of the list, the participant is presented with 30 grocery items (i.e., immediate recognition). The examinee should choose the ten target items (i.e., create the shopping list) from an extensive array of items, which also contains five qualitative distractors (e.g., semi-skimmed milk versus skimmed milk), five quantitative distractors (e.g., 1 kg potatoes versus 2 kg potatoes), and ten false items (e.g., bread, bananas etc.). Similarly, the participant should choose the same 10 items when they arrive at the supermarket approximately 20 minutes later (delayed recognition). The examinee gains 2 points for each correctly chosen item, 1 point for selecting a qualitative or quantitative distractor, and 0 points for the false items. Scores range from 0 to 20.

*Multitasking/Task-Shifting*. The ability to multitask is examined using a cooking task, where the participant should prepare and serve his or her breakfast (e.g., sausages, omelette, and a cup of tea/coffee) and place a chocolate pie in the oven. Scoring relies on the animations from each game object (i.e., the omelette and the sausages). At the beginning of the animation, both items have a reddish (raw) colour which gradually turns to either a yellowish (omelette) or brownish (sausages) colour, and finally both turn to black (burnt). The score for each pan depends on the time that the examinee removes the pan from the stove and places the pan on the kitchen worktop. Equally, the score for boiling the kettle is measured in relation to the stage of the audio playback (e.g., the kettle whistles when the water is ready) when the kettle is placed on the kitchen worktop. Scores range from 0 to 16.

*Planning.* Planning ability is assessed by asking participants to draw their route around the city (e.g., visiting the bakery, supermarket, library, and returning home) on a 3D interactive board. The road system comprises 23 street units. When the examinee selects a unit, 1 point is awarded. The ideal route to visit all three destinations is 15 units; hence, any extra or missing units are subtracted from the total possible score of 15. Up to 4 more points are awarded for the time taken to complete the task. Scores range from 0 to 19.

*Visuospatial attention.* Visuospatial attention is assessed by asking the participant to find and collect 6 specific items (i.e., a mobile phone, a £50 note, a library card, the flat keys, a red book, and car keys) from the living-room. A reminder of these items remains on the wall (i.e., the items are displayed as 3D objects with labels). However, there are also distractors (i.e., magazines, books, a remote control, a notebook, a pencil, a chessboard, and a bottle of wine) in the room. The examinee receives 2 point for each target item collected (6 target items). Also, up to 4 points are awarded for the speed of detecting the items. If the examinee attempts to collect one of the distractors, it counts as an error. Up to 4 points are awarded for the accuracy of detecting items. Scores range from 0 to 20.

*Visual attention.* Visual attention is measured while the participant is seated as a passenger in a car next to a driver. The participant should identify all the targets (i.e., 16 posters of a radio station) on both sides of the road, while s/he needs to avoid any distractors (i.e., 8 posters that are a different shape and 8 posters with a different background colour). The examinee is awarded 1 point when a target poster is "spotted" and subtracted 1 point when a distractor poster is "spotted". Scores range from 0 to 16.

*Auditory Attention.* Auditory attention is also examined while the participant is seated as a passenger next to a driver. The participant should detect all the target sounds (i.e., 16 bell sounds) presented on both sides of the road, while avoiding the distractor sounds (i.e., 8 high pitch bells, and 8 dongs). If the examinee presses the trigger on the right controller to detect a target sound originating on the right side (i.e., controller and sound on the same side), then s/he gets 2 points. If the examinee presses the trigger on the right controller to detect a target sound originating on the left side (i.e., controller on the opposite side), s/he gains only 1 point. If the examinee responds to a distractor sound, irrelevant of its origin or the controller used to respond, 1 point is deducted. Scores range from 0 to 32.

### Statistical Analyses

All the statistical analyses were conducted using R language (R Core Team, 2020) in RStudio (RStudio Team, 2020). The stats (R Core Team, 2020) and the car (Fox & Weisberg,

2019) packages were used for conducting the linear regression analyses and the diagnostics. The pwr package (Champely, 2020) was used to perform the statistical power analysis for each model. The ggstatsplot package (Patil, 2018) was used for visualising the results.

*Linear regression analyses*

The normality assumption for the event- and time-based scores was examined using the Shapiro-Wilk Normality test. The Non-Constant Error Variance test was used to check the homoscedasticity assumption for the models. The variance inflation factor for each model's predictors was calculated and checked to detect multicollinearity. A statistical power analysis was conducted for each model. Linear regression analyses were performed to investigate which everyday cognitive functions predict everyday PM functioning. The models were compared by conducting analyses of variance. The criteria for the comparisons between the models were the Akaike's information criterion (AIC), the F statistic and its significance level, the variance explained (i.e., $R^2$), and the statistical power of each model.

$1^{st}$ *step.* Initially, 9 separate models were created, each with a single cognitive predictor. The 9 models were compared against one another. The best model among them was then used to build the models in the next step.

$2^{nd}$ *step*. In this step, another 8 models were created, each with two cognitive predictors. In each model, the first predictor was the predictor from the best model in the previous step. The other predictor was one of the other cognitive function scores. These 8 models were then compared against each other. The best model was then compared to the best model from the previous step. If the best model from the current step was better than the best model from the previous step, then the predictors of the best model from the current step were used to build the models in the next step.

*Following steps.* The procedure described above was continued until the best model from the previous step was better than the best model from the current step. Then, the best model from the previous step was accepted as the best final model.

**Results**

The majority of participants reported no cybersickness symptoms (i.e., nausea, dizziness, disorientation, fatigue, or instability), while no one reported cybersickness symptoms stronger than a mild feeling, which postulates that cybersickness symptoms did not affect the participants' performance. The descriptive statistics for the VR-EAL scores are displayed in Table 2. The Shapiro-Wilk Normality test yielded non-significant results for both the event-based (p = .106) and time-based (p =.179) scores, postulating that the event- and time-based scores were normally distributed. Equally, the results of the Non-Constant Error Variance test for each model were not significant, indicating that each model's residuals were homogeneous. The variance inflation factors for the models' predictors varied from 1 – 1.6, suggesting an absence of multicollinearity.

Table 2. Descriptive statistics for the VR-EAL scores

|  | N | Mean (SD) | Range |
|---|---|---|---|
| Event-based (max = 24) | 41 | 18.15 (3.26) | 8-24 |
| Time-based (max = 18) | 41 | 11.63 (3.10) | 6-18 |
| Immediate Recognition (max = 20) | 41 | 15.51 (1.98) | 10-18 |
| Delayed Recognition (max = 20) | 41 | 17.17 (2.42) | 12-20 |
| Visual Attention Accuracy (max = 32) | 41 | 22.98 (3.84) | 17-30 |
| Visual Attention Speed (max = 32) | 41 | 23.61 (3.69) | 18-30 |
| Visuospatial Attention Speed (max = 16) | 41 | 11.90 (1.50) | 8-14 |
| Visuospatial Attention Accuracy (max = 16) | 41 | 12.10 (1.18) | 8-13 |
| Auditory Attention (max = 32) | 41 | 29.56 (3.66) | 20-32 |
| Planning (max = 19) | 41 | 14.90 (1.51) | 11-17 |
| Cooking Task (max = 16) | 41 | 9.68 (2.57) | 2-13 |

*Event-based score - linear regression models*

The models with a single predictor are displayed in Table 3. The delayed recognition model best predicted event-based PM functioning. The model explained 17% of event-based PM functioning and had adequate statistical power. Visuospatial speed also significantly predicted everyday event-based PM functioning and the model explained 11% of the variance. However, the model had insufficient statistical power.

Table 3. Event-based prospective memory models with a single predictor

| Model | F (1,39) | t (39) | p (>|t|) | β coefficient | AIC | $R^2$ | SP |
|---|---|---|---|---|---|---|---|
| Immediate Recognition | 3.29 | 1.82 | p = .078 | 0.46 | 215 | .08 | 47% |
| Delayed Recognition | 7.99 | 2.83 | p = .007** | 0.56 | 211 | .17 | 82% |
| Visual Accuracy | 2.14 | 1.46 | p = .151 | 0.19 | 216 | .05 | 29% |
| Visual Speed | 2.24 | 1.50 | p = .142 | 0.21 | 216 | .05 | 29% |
| Visuospatial Accuracy | 0.54 | 0.74 | p = .466 | 1.78 | 218 | .01 | 10% |
| Visuospatial Speed | 4.63 | 2.15 | p = .038* | 0.71 | 214 | .11 | 58% |
| Auditory Attention | 3.00 | 1.73 | p = .091 | 0.24 | 215 | .07 | 42% |
| Planning | 2.52 | 1.59 | p = .120 | 0.53 | 216 | .06 | 34% |
| Multitasking | 1.72 | 1.31 | p = .198 | 0.26 | 217 | .04 | 24% |

AIC = Akaike's information criterion; A smaller AIC indicates a better fit for the model.
($R^2$ x 100) indicates the percentage of variance explained by the model.
SP = statistical power. β coefficient = standardised regression coefficient.

After several steps, the best model for event-based PM functioning is visualised in Figure 3. Based on the F statistic and its p-value, the combining effect of delayed recognition, visuospatial speed, and planning ability on everyday event-based PM functioning was found to be significant. The model explained 25% of the variance, and it had substantial statistical power. Based on the standardised regression coefficients, delayed recognition, visuospatial speed, and planning ability predict everyday event-based PM functioning. However, only the effect of delayed recognition was found to be statistically significant.

Figure 3. Best model for event-based prospective memory

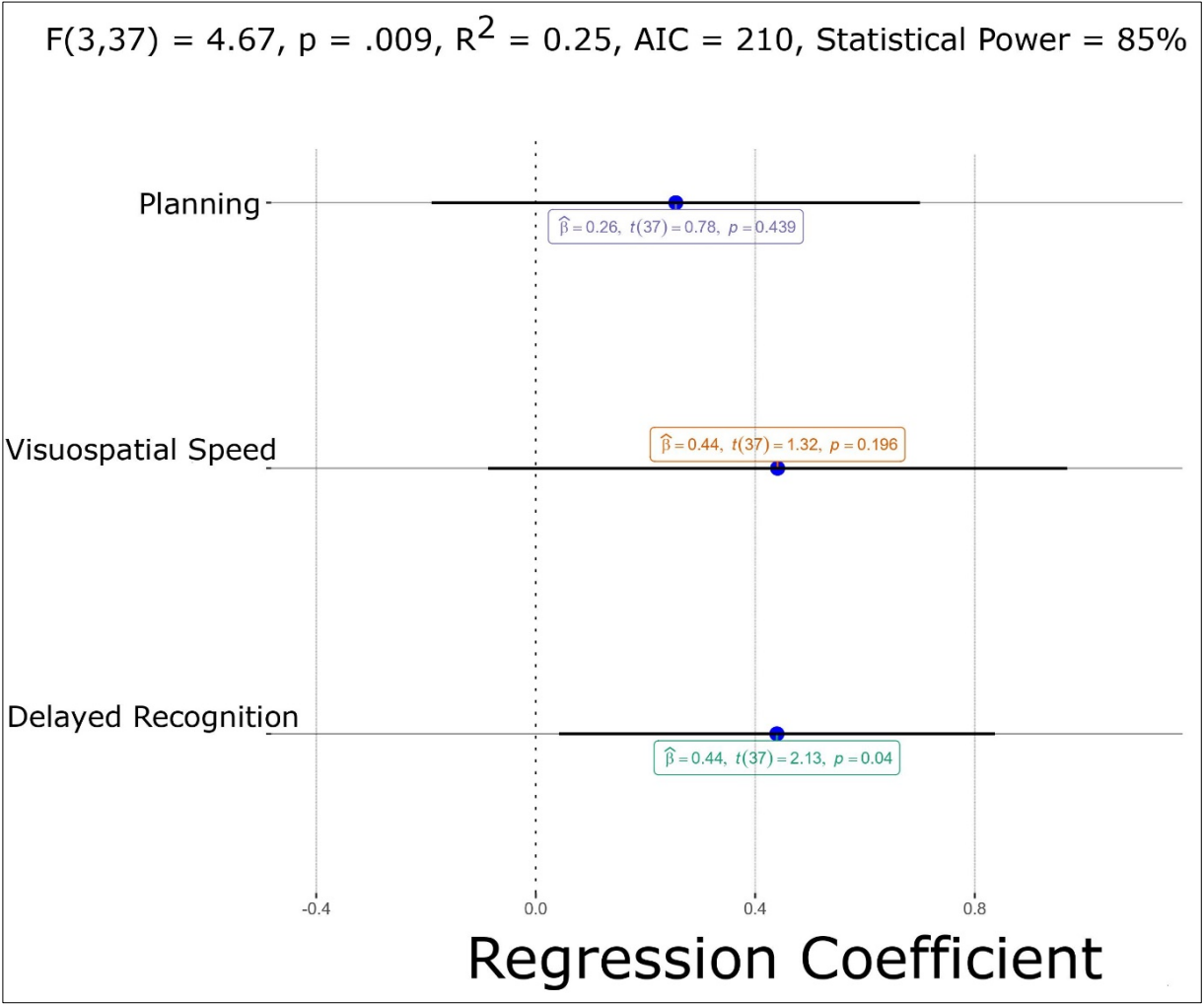

*Time-based score - linear regression models*

The models with a single predictor are displayed in Table 4. The model with planning ability was found to best predict time-based PM functioning. The model explained 12% of the variance. However, its statistical power was insufficient. Moreover, the effect of visuospatial accuracy on time-based PM functioning was marginally insignificant. The model appeared to explain only the 9% of the variance of time-based PM functioning, and its statistical power was also inadequate.

Table 4. Time-based prospective memory models with a single predictor

| Model | F (1,39) | t (39) | p (>|t|) | β coefficient | AIC | $R^2$ | SP |
|---|---|---|---|---|---|---|---|
| Immediate Recognition | 0.16 | 0.40 | p = .690 | 0.10 | 214 | .01 | 10% |
| Delayed Recognition | 2.36 | 1.54 | p = .132 | 0.31 | 212 | .06 | 34% |
| Visual Accuracy | 0.86 | 0.93 | p = .361 | 0.12 | 213 | .02 | 14% |
| Visual Speed | 1.00 | 1.00 | p = .325 | 0.13 | 213 | .03 | 19% |
| Visuospatial Accuracy | 3.72 | 1.93 | p = .061 | 4.28 | 210 | .09 | 51% |
| Visuospatial Speed | 0.69 | 0.83 | p = .410 | 0.27 | 213 | .02 | 14% |
| Auditory Attention | 2.80 | 1.67 | p = .102 | 0.22 | 211 | .07 | 42% |
| Planning | 5.06 | 2.25 | p = .030* | 0.69 | 209 | .12 | 65% |
| Multitasking | 1.21 | 1.10 | p = .279 | 0.21 | 213 | .03 | 19% |

AIC = Akaike's information criterion; A smaller AIC indicates a better fit for the model.
($R^2$ x 100) indicates the percentage of variance explained by the model.
SP = statistical power.  β coefficient = standardised regression coefficient.

The best final model for time-based PM functioning is illustrated in Figure 4. The F statistic and its p-value indicate that the combined effect of planning, visuospatial accuracy, delayed recognition, and multitasking/task-shifting abilities on everyday time-based PM functioning is significant. The model explained 32% of the variance and it had a substantially high statistical power. The standardised regression coefficients indicated that planning, visuospatial accuracy, delayed recognition, and multitasking/task-shifting predict everyday time-based PM functioning. However, only the effects of planning and visuospatial accuracy were found to be statistically significant.

Figure 4. Best model for time-based prospective memory

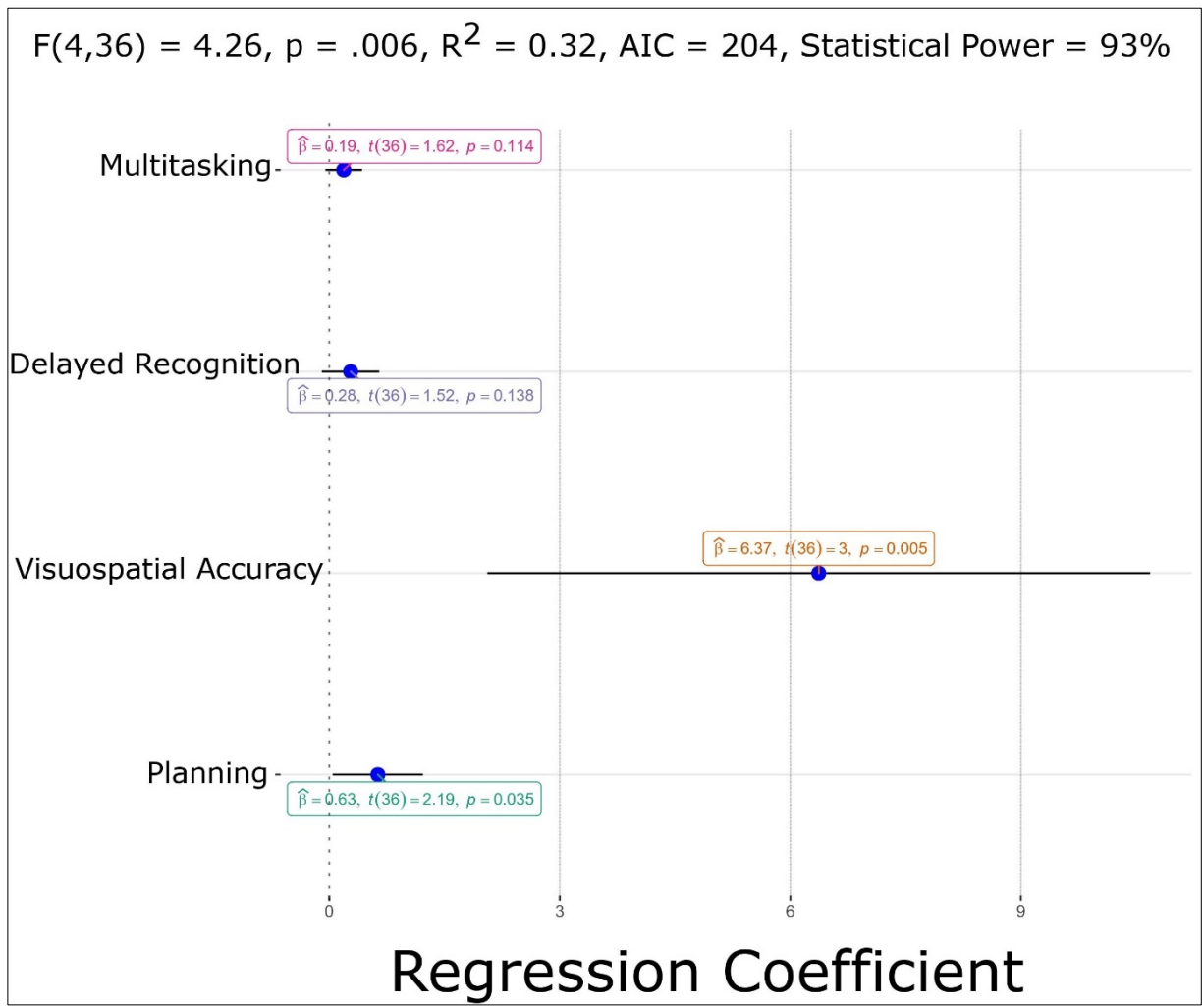

**Discussion**

This study aimed to examine the cognitive functions implicated in everyday PM functioning using immersive VR. The findings show that delayed recognition and planning ability significantly predict event- and time-based PM functioning respectively. The final models included delayed recognition, visuospatial attention speed, and planning ability for event-based PM functioning and planning, visuospatial attention accuracy, delayed recognition, and multitasking/task-shifting ability for time-based PM functioning. These findings highlight the importance of attention, memory, and executive functioning in everyday PM functioning.

*Event-based prospective memory*

Delayed recognition explained some of the variance in event-based everyday PM functioning. This is in line with previous lab-based studies that have demonstrated that memory processes influence PM functioning (e.g., Smith, 2003; Smith et al., 2007; McDaniel & Einstein, 2000, 2007). The current findings indicate similarities with episodic memory, where better memory encoding and maintenance (i.e., in this case the PM intended intention) facilitates more efficient retrieval (Frankland, Josselyn, & Köhler, 2019; Tulving, 1983; Tulving & Thomson, 1973; Xue, 2018). In PM, the association between an X PM cue and Y PM action appear to facilitate better encoding as well as maintenance of the intended PM action (Gonneaud *et al.*, 2011; McDaniel & Scullin, 2010; Zuber et al., 2019), which highlights the role of delayed recognition in everyday event-based PM functioning. Indeed, Gonneaud and collaborators (2011) found that retrospective processes are significant in event-based PM functioning, but they are of secondary importance in time-based PM.

Furthermore, the inclusion of visuospatial attention speed (i.e., the ability to select and scan an area of interest to detect an item of interest) in our event-based PM model emphasizes the importance of contextual information in PM functioning. Again, this supports previous findings (e.g., Marsh, et al., 2008; Scullin *et al.*, 2013; Shelton & Scullin, 2017). It is also in line with the dynamic multiprocess framework, where attentional processes are implemented for contextual information and/or the plan of action (Scullin *et al.*, 2013; Shelton & Scullin, 2017). In our best model for event-based PM functioning, both visuospatial attention speed and planning ability were included. Visuospatial attention speed may indicate the ability to rapidly scan the surrounding environment to retrieve information, which may be valuable for informing and/or updating the formulated plan of action. In line with our findings, the findings of several studies have postulated the role of planning in event-based PM (e.g., Gonneaud *et al.*, 2011; McDaniel & Scullin, 2010; Zuber, Kliegel, & Ihle, 2016; Zuber et al., 2019).

*Time-based prospective memory*

Planning ability was found to explain a substantial amount of the variance in time-based PM functioning. This highlights the important role of planning ability in time-based PM functioning, which is in line with previous studies that have also demonstrated that planning influences time-based PM (e.g., Azzopardi et al., 2017; Liu & Park, 2004; McFarland & Glisky, 2009; Mioni & Stablum, 2014; Vanneste et al., 2016). Beyond formulating an effective plan of action, the inclusion of planning may also indicate the importance of strategic time monitoring, which avoids the cognitively demanding task of constantly time monitoring (Cona *et al.*, 2012; Kliegel *et al.*, 2001). In strategic time monitoring, when individuals check the time, they estimate when they should check it again based on a predefined plan of action, and adjust correspondingly the frequency of time monitoring (Cona *et al.*, 2012; Kliegel *et al.*, 2001).

Similarly, the involvement of visuospatial attention accuracy in time-based PM may be an indication of effective time monitoring. In VR-EAL, the watch is attached to the left hand/controller of the participant which is accessible at all stages of the scenario (see scene 10 in Figure 1). Most of the time, the watch is not within the participant's view (see Figures 1 and 2). Time monitoring requires the participant to detect and focus on the watch. Consequently, visuospatial attention accuracy highlights the crucial role of time monitoring in time-based PM functioning (McFarland & Glisky, 2009; Mioni & Stablum, 2014; Vanneste et al., 2016). However, this finding may also indicate the importance of contextual information in time-based PM functioning. For example, the detection of a medication bottle may remind the individual that s/he needs to take her/his medicine at a specific time. This contextual information may assist the individual in implementing strategic time monitoring, as well as maintaining the PM intention.

The delayed recognition was also included in the time-based PM model, which supports previous findings, where episodic memory was implicated in time-based PM functioning (e.g., Mackinlay, Kliegel, & Mäntylä, 2009; McFarland & Glisky, 2009). However, the effect of delayed recognition was not significant, which indicates that its role in our model appears to be supplementary, as was previously seen in Gonneaud and collaborators (2011). Finally, the inclusion of task-shifting in the best model of time-based PM appears to be in accordance with the findings (e.g., Azzopardi *et al.*, 2017; Gonneaud *et al.*, 2011; McFarland & Glisky, 2009; Mioni & Stablum, 2014; Vanneste *et al.*, 2016; Zuber *et al.*, 2019). However, in our study, task-shifting did not appear to have an important effect on time-based PM functioning. Its inclusion in the best model may have a secondary role, and indicate the ability to inhibit the task at hand, and shift attention towards time monitoring.

### *Ecological Validity*

As previously suggested, the implementation of lab-based, non-ecological PM paradigms may provide findings which are discrepant with the findings of ecologically valid research paradigms (Marsh et al., 1998). This has been further supported by the so-called age-prospective memory-paradox where significant age-related differences were found on laboratory PM tasks, while they were not found with ecologically valid PM tasks (Kvavilashvili *et al.*, 2013; Niedźwieńska & Barzykowski, 2012; Phillips *et al.,* 2008; Uttl, 2008). These discrepancies may be attributed to the use of compensatory strategies and external memory aids, which are common strategies in daily life for performing PM tasks (Phillips *et al.,* 2008; Uttl, 2008). Our findings further support the necessity for adopting ecologically valid paradigms to study everyday PM functioning.

In our current study, the implementation of a long-lasting scenario (i.e., approximately 70 minutes) demonstrated the importance of recognising a cue after a long delay (i.e., delayed recognition), especially for event-based PM. Importantly, the VR-EAL attempts to simulate a common day in the real-world. The inclusion of several everyday tasks (e.g., preparing breakfast, shopping, finding the keys, transporting, interacting with others, and taking meds) highlighted the importance of prioritising the activities (i.e., planning ability) in everyday PM functioning, especially for time-based PM. Furthermore, the implementation of a complex, dynamic, and immersive 360º environment appeared to highlight the importance of visuospatial attention processes in everyday PM functioning, while the simple visual attention processes did not appear crucial. Lastly, enabling participants to perform the VR-EAL tasks through ergonomic and naturalistic ways seemed to diminish the importance of multitasking/task-shifting ability in PM functioning, as has been previously seen in laboratory paradigms. Our findings, hence, postulate the importance of ecological validity for examining PM functioning and generalising the findings to daily life. Also, our findings suggest the importance of cognitive abilities in everyday PM functioning, such as episodic memory, visuospatial attention, and planning, which were either undermined (e.g., delayed recognition and planning) or not even considered (e.g., visuospatial attention) in non-ecologically valid laboratory paradigms, where the duration is short (e.g., 5-15 minutes), there is only a single or small number of tasks, and the testing environment is simple and limited (e.g., 90 - 180º field of view). The role and importance of these cognitive functions should be further studied in future ecologically valid studies.

*Limitations and future studies*

Our study has some limitations. The sample size was relatively small, although it allowed us to perform sound statistical analyses with high statistical power. Future studies should include a larger population which would allow a factorial analysis involving the components of PM. Also, our participants were only young adults. A more diverse population including both younger and older adults may help to elucidate the effect of aging on PM assessed using the VR-EAL. VR-EAL also has the limitation that it cannot measure time monitoring and this should be integrated into a future version of the VR-EAL. Furthermore, while simulations of everyday life like the VR-EAL are suitable for the investigation of real-life PM, they may be susceptible to confounding factors, such as environmental stimuli (e.g., a building with several signs) or an interaction with a 3D character (i.e., the requirement of communicational skills), which were not experimentally controlled. Laboratory tasks have the advantage of isolating confounding factors and allowing the examination of specific processes. However, laboratory PM tasks in their current form suffer from limitations such as a two-dimensional environment, keyboard-based responses, static stimuli, and a lack of realism. A solution would be the modernization of laboratory tasks using immersive VR technology, which would reduce the divergence from real-life conditions, and enable a meticulous examination of PM and the role of other cognitive functions in everyday PM functioning.

*Conclusions*

The findings of this study demonstrated the importance of delayed recognition, planning, and visuospatial attention in everyday PM. Delayed recognition and planning ability were found to be especially crucial in event- and time-based PM respectively. In order of importance, delayed recognition, visuospatial attention speed, and planning ability were found to be involved in event-based PM functioning. Comparably, planning, visuospatial attention accuracy, delayed recognition, and multitasking/task-shifting ability were found to be involved in time-based PM functioning. These findings further suggest the importance of ecological validity in the study of PM, which may be achieved using immersive VR paradigms.

**Acknowledgments**

The authors declare no conflicts of interest. The current study did not receive any financial support or grants.